\documentclass[aps, pra, showpacs, amsmath, twocolumn, groupedaddress]{revtex4}

\usepackage{graphics}

\newcommand{\rhored}{\rho^\mathrm{red}}

\renewcommand{\vec}[1]{\ensuremath{\boldsymbol{#1}}}
\newcommand{\hilbert}{\ensuremath{\mathcal{H}}}
\newcommand{\Bra}[1]{\left\langle{#1}\right\rvert}
\newcommand{\Ket}[1]{\left\lvert{#1}\right\rangle}
\newcommand{\Jm}[2]{\begin{smallmatrix}{#1}\\{#2}\end{smallmatrix}}
\newcommand{\Jmjm}[4]{\begin{smallmatrix}{#1}&{#3}\\{#2}&{#4}\end{smallmatrix}}

\newcommand{\Braket}[1]{\left\langle{#1}\right\rangle}
\newcommand{\largeBraket}[3]{%
 \Bigl\langle{#1}\Bigr\lvert{#2}\Bigl\lvert{#3}\Bigr\rangle}
\newcommand{\centerBraket}[3]{%
 \left\langle{#1}\left\lvert{#2}\right\lvert{#3}\right\rangle}
\newcommand{\cgBraket}[2]{\Bigl\langle{#1}\Big{|}{#2}\Bigr\rangle}
\newcommand{\smashcgBraket}[2]{\bigl\langle{#1}\big{|}{#2}\bigr\rangle}
\newcommand{\smashKet}[1]{\bigl\lvert{#1}\bigr\rangle}

\begin{document}
\title{Mixed collective states of many spins}
\author{Janus Wesenberg}
\author{Klaus M{\o}lmer}
%\thanks{}
\affiliation{
%  Quantop (Danish Quantum Optics Center),
  QUANTOP,
  Danish Quantum Optics Center,
  Institute of Physics and Astronomy, 
  University of Aarhus,
  DK-8000 {\AA}rhus C, 
  Denmark}

\date{\today}

\begin{abstract}
Mixed states of samples of spin $s$ particles which are symmetric under
permutations of the particles are described in terms of their
total collective spin quantum numbers. 
We use this description to analyze the influence on spin squeezing due
to imperfect initial state preparation.
\end{abstract}

%  03.67.-a  Quantum information
%  42.50.-p  Quantum optics
% \keywords{}

\maketitle
% \section{\label{}}

\section{Introduction}
Atomic samples for which external perturbations and mutual
interactions between the atoms are completely symmetric under
permutations of the atoms are conveniently described in terms of
collective variables.  
In particular, if the atoms have two internal
states relevant for the dynamical processes, each atom can be
represented as a spin $\frac{1}{2}$ particle, and the full Hamiltonian of the
system can be expressed in terms of the collective spin operator
$\vec{J}=\sum_i \vec{S}_i$. 
A state vector of the system which reflects the permutation
symmetry of $n$ two-level particles can be expanded on the set
of symmetric Dicke states $\Ket{\Jm{j_0}{m}}$, where $j_0 = n/2$, and
the  quantum number $m$ ranges from $-j_0$, where all the
atoms populate the states associated with a spin down single particle
state, to $j_0$ where all are in the spin up state. Intermediate
values represent states containing all possible permutations of
the atoms populating either of the two spin $\frac{1}{2}$ components,
but with a definite total number, $j_0\pm m$, of atoms in the
spin up ($+$) and down states ($-$).

Collective behavior of atomic ensembles is at the root of
superradiance, and the Dicke states were introduced to account for
atom-field interactions already many years ago.  The past decade has
provided numerous discussions and analyses of collective atomic
dynamics in connection with atomic clocks and quantum information
\cite{WB94,NC00},
and recently a number of promising schemes
has been presented for tailoring of the collective atomic spin noise,
relying on atom-light interaction 
\cite{kuzmich97:spin_squeez_ensem_atoms_illum, hald99:spin_squeez_atoms},
collisional interactions in atomic samples 
\cite{duan00:squeez_entan_atomic_beams,soerensen01:many_partic_entan_with_bose,
xBM01}, 
and on quantum non-demolition measurements 
\cite{KM00,JK01}.

Restricting calculations to the space spanned by the symmetric Dicke
states implies a formidable simplification of the full $2^n$
dimensional problem to an $n+1$ dimensional one, and use of algebraic
properties of the angular momentum operators further facilitate
calculations of properties of the physical system.  A large number of
theoretical analyses therefore presents calculations restricted to
this subspace.
But, even if the assumption holds that the systems is invariant under
exchange of atoms, not only the symmetric Dicke states will be
populated: Imagine two two-level atoms, both described by the mixed
state density matrix
\begin{equation}
  \rho_1 =
  p \Ket{\uparrow}\Bra{\uparrow} + (1-p) \Ket{\downarrow}\Bra{\downarrow}.
  \label{eq:rho_p}
\end{equation}
The combined state $\rho_2=\rho_1\otimes\rho_1$ of both atoms is also
a mixed state. It is manifestly symmetric under exchange of the atoms,
but if $p\neq 0$ or $1$ it is a mixture of all $2^2=4$ product states
spanning the full Hilbert space of the pair of particles, and by a
basis change we see that both the spin singlet and the three spin
triplet eigenstates of the collective spin operator are populated.
Unlike the spin singlet wave function, a density matrix component
proportional to a projector on the spin singlet state is, indeed,
invariant under exchange of the particles.

Symmetric mixed states incoherently populate state vectors of different
permutation symmetries.  It is the purpose of the present paper to
characterize this population of states of lower symmetry and to
investigate the effect, {\it e.g.}, on the achievements of spin squeezing
and multi-particle entanglement of samples of atoms. For
experiments, the situation with values of $p$ close to zero or unity
are particularly interesting as they represent the degree
of polarization obtained, {\it e.g.}, by optical pumping.

In section \ref{sec:expect-valu-coll} we derive some expectation
values for the collective spin operators.
In section \ref{sec:dens-matr-tens}, we derive the probability
distribution for states of different $j$ and $m$, $j \leq j_0$, for a
collection of spin $\frac{1}{2}$ particles in the state \eqref{eq:rho_p}.
In section \ref{sec:effect-spin-sque}, we analyze the
achievements of spin squeezing on such a collection of particles.
In section \ref{sec:atoms-with-more} we discuss the case of
$s>\frac{1}{2}$, 
{\it i.e.}, atoms with more than two states.

\section{\label{sec:expect-valu-coll}Expectation values of the collective spin operators}
Before determining the full density matrix of a system of uncorrelated
atoms, it is instructive to see how much information can be derived
from the fact that the atoms are uncorrelated.

For a system of $n$ uncorrelated atoms, we may determine the variances
of the collective spin operators through
\begin{equation}
  \label{eq:vartruth}
  \Delta J_\alpha^2 =n \Delta S_\alpha^2 \quad \alpha=x,y,z.
\end{equation}
In the case where the individual atoms are two-level atoms distributed
according to \eqref{eq:rho_p}, we obviously have
$\Braket{\vec{S}^2}=\frac{3}{4}$ and
$\Braket{S_z^2}=\frac{1}{4}p+\frac{1}{4}(1-p)=\frac{1}{4}$,
assuming $\hbar=1$.
Since the distribution is invariant under rotations around the
$z$-axis, the $x$- and $y$-variances must be equal and we find
\begin{equation}
  \label{eq:varxy}
  \Delta J_x^2 =\Delta J_y^2
  =\frac{n}{2} \Braket{\vec{S}^2-S_z^2}=\frac{1}{2}\, j_0.
\end{equation}
Also, since the expectation value of $S_z$ is $p-\frac{1}{2}$,
the variance of $J_z$ is
\begin{equation}
  \label{eq:varz}
  \Delta J_z^2
   =n \left( \Braket{S_z^2}-\Braket{S_z}^2 \right) 
           = p(1-p)\,2 j_0. 
\end{equation}

To find $\Braket{\vec{J}^2}$ we may again utilize the fact that the atoms
are independent so that
$\Braket{\vec{J}^2} = \sum_{i\neq j} \Braket{\vec{S}_i}\cdot\Braket{\vec{S}_j}
+\sum_i \Braket{\vec{S}_i^2}$,
which in this case simplifies to 
\begin{equation}
  \label{eq:j2expb}
  \Braket{\vec{J}^2} = (2p-1)^2 j_0(j_0+1) + 6 p (1-p) j_0.  
\end{equation}
In the completely polarized case, $p=1$, the total spin equals
$j_0=n/2$ and $\Braket{\vec{J}^2}=j_0(j_0+1)$.
For mixed ensembles $\Braket{\vec{J}^2}$ is reduced, reflecting a
distribution over states with different $j$, and in the completely
unpolarized case, $p=\frac{1}{2}$, each atom contributes $\frac{3}{4}$
to $\Braket{\vec{J}^2}=n\, \frac{3}{4}$. 
In the following section we shall determine the population of states
with different $j$ for any value of $p$.

\section{\label{sec:dens-matr-tens}The density matrix for a tensor product of mixed states}
$n$ spin $\frac{1}{2}$ particles span a $2^n$ dimensional Hilbert
space, $\hilbert$, with basis vectors $\Ket{m_1, ... ,m_n}$, 
$m_i=\pm \frac{1}{2}$.
An alternative representation is by means of eigenstates,
$\Ket{\Jm{j}{m};\lambda}$, of the collective spin operators 
$\vec{J}^2$ (eigenvalue $j(j+1)$) and $J_z$ (eigenvalue $m$), where
$\lambda$ enumerates the degenerate states within the eigenspaces,
$\hilbert_{j,m}$, of the $\vec{J}^2$ and $J_z$ operators.

\subsection{\label{sec:coll-spin-stat}Collective spin states for $n$ 
spin $\frac{1}{2}$ particles}
A complete description of the full set of basis states is given in
Ref.~\cite{AC72}, but for completeness we shall derive our results
by a formally simpler analysis. 

The $\Ket{\Jm{j}{m};\lambda}$ states
must span the full $2^n$ dimensional Hilbert space, and we start out
by identifying the number of states with a given value of the quantum
number $m$. Since $m=\sum_i m_i$, the total number of atoms in the
spin up state $\Ket{\uparrow}$ is $j_0+m$, and these states can be
selected in a number of ways given by the combinatorial factor
\begin{equation}
N_m= \binom{n}{j_0+m} = \frac{n!}{(j_0+m)!(j_0-m)!}.
\end{equation}
Due to the binomial formula $(1+1)^n=\sum_k \binom{n}{k} 1^k 1^{(n-k)}$
we verify that $\sum_m N_m=2^n$.

States with given $m$ values should now be grouped according to the
values of $j$. Clearly $j \geq m$, hence we may begin with $m=j_0$ for
which only a single state exists, and for which $j$ must equal $j_0$.
The combinatorial factor $N_{m=j_0-1}$ equals $n$ (a single spin down
particle can be chosen in $n$ different ways), and since only one of
the states with $m=j_0-1$ can be obtained by the collective lowering
operator $J_-$ acting on the state
%$\Ket{\Jm{j_0}{j_0}}$, 
$\smashKet{\Jm{j_0}{j_0}}$, 
the remaining $n-1$
states must have $j=j_0-1$. Applying the lowering operator $J_-$ on
all the identified $m=j_0-1$ states produces one state with $j=j_0$
and $n-1$ states with $j=j_0-1$. The remaining $N_{m=j_0-2}-n$ must
therefore have $j=j_0-2$. Continuing this argument provides the total
number of states with $j\geq m$ for any $m\geq 0$, and by symmetry we
obtain the same number of $j$-states for negative $m$.  In conclusion,
the number of states with given values of $j$ and $m$ is independent
of $m$ and equal to $D^{(1/2)}_j=N_{m=j}-N_{m=j+1}$:
\begin{equation}
 \dim \hilbert_{j,m}=
 D^{(1/2)}_j=\frac{2j+1}{j_0+j+1} \binom{2 j_0}{j_0+j}.
\label{eq:SM_dimension}
\end{equation}

\subsection{\label{sec:dens-matr-tens-1}The density matrix for a tensor product of mixed states}
Consider now the n-particle density operator
\begin{equation}
\rho_n = \rho_1^{\otimes n},
\label{eq:rho_N}
\end{equation}
{\it i.e.}, the tensor product of $n$ identical two-state density
matrices. 
Since any two-state density matrix may be diagonalized through
rotations, we will only consider the diagonal form of $\rho_1$ as
given by \eqref{eq:rho_p}.

As we intend to demonstrate the usefulness of the collective spin
operator, we will only be interested in the reduced density matrix,
$\rhored_n$, obtained by taking the partial trace over the
remaining degrees of freedom:
\begin{equation}
  \label{eq:rhored}
  \largeBraket{\Jm{j'}{m'}}{\rhored_n}{\Jm{j}{m}}\equiv  
  \sum_\lambda 
  \largeBraket{\Jm{j'}{m'};\lambda}{\rho_n}{\Jm{j}{m};\lambda}.
\end{equation}

To determine the $p$-dependence of the matrix elements of 
$\rhored_n$, we note that the vector
$\Ket{\Jm{j}{m};\lambda}$ may be expanded on the product basis, 
$\left\{\Ket{m_1,m_2,\ldots,m_n}\right\}$, so that \eqref{eq:rhored}
only involves matrix elements of the form
\begin{equation}
  \label{eq:melemform}
  \centerBraket{m'_1,m'_2,\ldots,m'_n}{\rho_1^{\otimes n}}{m_1,m_2,\ldots,m_n},
\end{equation}
where $\sum m'_i=m'$ and $\sum m_i=m$.
Since $\rho_1$ is diagonal, only matrix elements with $m'_i=m_i$ are
non-vanishing, and consequently $\rhored_n$ is diagonal in $m$.
Writing the diagonal elements of $\rho_1$ as
$p^{1/2+m_s}\,(1-p)^{1/2-m_s}$,
it is evident that the value of any non-vanishing matrix element of the
form \eqref{eq:melemform} is $p^{j_0+m} (1-p)^{j_0-m}$.  Since the
expansions coefficients used to write $\rhored_n$ in terms of
these matrix elements are independent of $p$, we conclude that this is the
full $p$-dependence of the matrix elements of $\rhored_n$.

We now consider the case where $p=\frac{1}{2}$, so that the single
particle density matrix, $\rho_1$, is proportional to the identity.
This implies that the full density matrix \eqref{eq:rho_N} is also
proportional to the identity, or to be more specific $\rho_n = 1/\dim
\hilbert$ in \emph{any basis}.
Consequently, since the entries of $\rhored_n$ are obtained by
performing a partial trace as defined in \eqref{eq:rhored},
$\rhored_n$ is diagonal, and the values of the diagonal
elements are determined by the dimension of the relevant $j,m$
subspaces, $\hilbert_{j,m}$.
Denoting the diagonal elements (populations) of $\rhored_n$ by
$p^{(n)}_{j,m}$, and using the dimension of $\hilbert_{j,m}$ found in 
\eqref{eq:SM_dimension}, this gives us
\begin{equation}
  \label{eq:rhoarg1}
  p^{(n)}_{j,m}=
  \frac{\dim{\hilbert_{j,m}}}{\dim{\hilbert}} =
  2^{-n} D^{(1/2)}_j \quad \text{for $p=\frac{1}{2}$}.
\end{equation}

Finally, for this form of $\rhored_n$ at $p=\frac{1}{2}$ to
agree with the general $p$-dependence as found above, we find that
$\rhored_n$ must be diagonal in $j$ for all values of $p$:
Assume to the contrary that an off-diagonal element exists which is
non-vanishing at some $p$-value. The required $p$-dependence implies
that such a matrix element would also be non-vanishing at
$p=\frac{1}{2}$ in contradiction with \eqref{eq:rhoarg1}.
Consequently such elements do not exist.

In conclusion, $\rhored_n$ is completely diagonal, and the
explicit form of the diagonal elements is
\begin{equation}
  \label{eq:rhojm}
  p^{(n)}_{j,m}=\frac{2j+1}{j_0+j+1} \binom{2 j_0}{j_0+j} p^{j_0+m} (1-p)^{j_0-m}.
\end{equation}

This surprisingly simple form of the full density matrix of an
ensemble of identically distributed two-level systems immediately
yields a host of interesting results.

As an important example we may use the result to illuminate the point
that the states with less than full permutation symmetry are
populated.  First, the marginal $j$-distribution, $p_j \equiv \sum_m
p^{(n)}_{j,m}$, is found to be
\begin{equation}
  \label{eq:margj}
  p_j =D^{(1/2)}_j
  \frac{p^{2j+1}-(1-p)^{2j+1}}{2p-1} (p(1-p))^{j_0-j}. 
\end{equation}
For $p=\frac{1}{2}$, $p_j=(2j+1) 2^{-n} D^{(1/2)}_j$ according to \eqref{eq:rhoarg1}.
For convenience we shall assume $p>\frac{1}{2}$ in the remainder of
this section, and consequently it seems justifiable to approximate 
\eqref{eq:margj} by dropping the term
$(1-p)^{2j+1}$ in the numerator, thus obtaining
\begin{equation}
  \label{eq:margjappr}
  p_j \approx \frac{2j+1}{j_0+j+1} \frac{p}{2p-1}
  \binom{2 j_0}{j_0+j} p^{j_0+j} (1-p)^{j_0-j}.
\end{equation}
We observe that the rightmost three terms constitute a normalized
binomial distribution with an expectation value for $j$ of
$j_c = (2p-1)j_0$ and a variance of   
$\sigma^2 = p (1-p) 2 j_0$.
To justify the approximation made above in a more rigorous way, we may
note that the omitted term corresponds to a similar binomial
distribution centered at $-j_c$. 
The approximation is thus expected to be valid as long as
$-j_c+\sigma < 0$, or equivalently 
$p-\frac{1}{2} > 1/\sqrt{8 j_0}$. 

\begin{figure}[htbp]
  \includegraphics{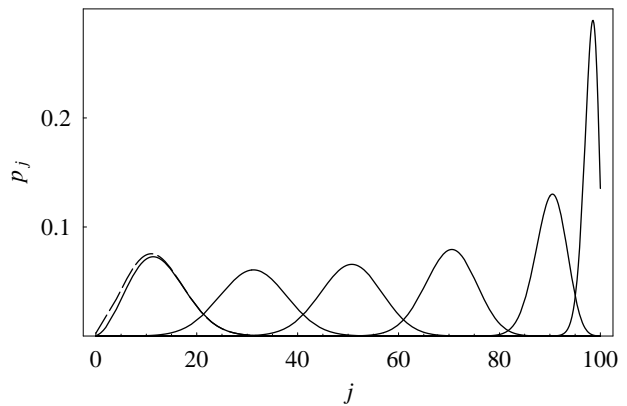}
  \caption{\label{fig:1}The distribution of the collective spin of $200$ spin $\frac{1}{2}$
    particles at various polarizations.  From left to right the peaks
    correspond to values of the polarization parameter $p$ in the
    single particle density matrix \eqref{eq:rho_p} of $0.535$,
    $0.65$, $0.75$, $0.85$, $0.95$ and $0.99$ respectively.  The solid
    line is the exact value of $p_j$ according to \eqref{eq:margj},
    the dashed line corresponds to the approximation
    \eqref{eq:margjappr} which cannot not be discerned from the exact
    values at large $p$-values. Note that the $p=0.99$ line ending in
    mid-air is not a technical artifact, but illustrates a significant
    population of the symmetric Dicke states as given by
    \eqref{eq:rhodicke}.}
\end{figure}

The front-factor in the approximation \eqref{eq:margjappr} is slowly
varying for $j$ not too close to $0$ and it is close to unity
at $j=j_c$, which suggests that we may describe the marginal $j$
distribution \eqref{eq:margj} by the binomial distribution alone.
That this is indeed a good picture is illustrated in Fig.~\ref{fig:1}, 
where we also note that the limit of validity deducted above 
for the approximation \eqref{eq:margjappr} seems to
be correct, as the only peak where the approximation deviates
visibly from the exact value corresponds to
$p=\frac{1}{2}+1/\sqrt{8 j_0}\approx 0.535$
chosen at the validity limit.

The variance in $j$ may be approximated by the variance of
the binomial distribution, $\sigma^2$, which according to 
\eqref{eq:varz} is equal to $\Delta J_z^2$.  
Since $\sigma^2$ decreases
with increasing $p$, we see that the $j$ states close to the maximal
value of $j_0$ are only populated at very high polarizations, in
agreement with Fig.~\ref{fig:1}.  In particular the total population of the
symmetric Dicke states ($j=j_0$) is given by 
\begin{equation}
  \label{eq:rhodicke}
  p_{j_0} = \frac{1}{2p-1} \left(p^{n+1}-(1-p)^{n+1}\right),
\end{equation}
according to \eqref{eq:rhojm}.
To have $p_{j_0} > p_{j_0-1}$ we must have $1-p < 1/n$, corresponding
to an expectation of finding less than one atom in the spin down state.
Note, however, that already at $1-p < \log 2/n$, we have half the
total population in the symmetric Dicke states.

\section{\label{sec:effect-spin-sque}The effect of spin squeezing on 
  a collection of spin $\frac{1}{2}$ particles} 
The variance of the spin components orthogonal to the mean spin of the
state \eqref{eq:rho_N} obeys 
$  \Delta J_x\, \Delta J_y = j_0/2 $, which exceeds the Heisenberg
uncertainty limit,
$\frac{1}{2}\left|\Braket{\left[J_x,J_y\right] }\right|$,
by a factor of 
$\left|2p-1\right|^{-1}$,
which approaches unity in the highly polarized cases ($p=0$, $1$) as
expected.  

It is interesting to study the effect of spin squeezing
operations on such states, since they might serve as models of
imperfectly prepared initial states on which squeezing operations are
performed.
Note that obtaining a spin squeezed state indicates that the
initially uncorrelated atoms described by \eqref{eq:rho_N} have become
entangled \cite{SM01}.

Spin squeezing may be performed in a number of ways, and we will
discuss the two unitary squeezing methods analyzed in
Ref.~\cite{KU93} in the case of initially pure distributions.
Both of these approaches make use of a Hamiltonian which is 
quadratic in collective spin variables, {\it i.e.}, it involves
interactions among the particles. The description of the 
evolution of the states would be immensely complicated in the 
product state representation of the many spins where, in fact, very
entangled states are produced. 
In the collective spin representation, however, the action
only has to be computed for each total spin component, which
is a much easier task.

\subsection{\label{sec:one-axis-twisting}One-axis twisting}
\begin{figure}[htbp]
  \includegraphics{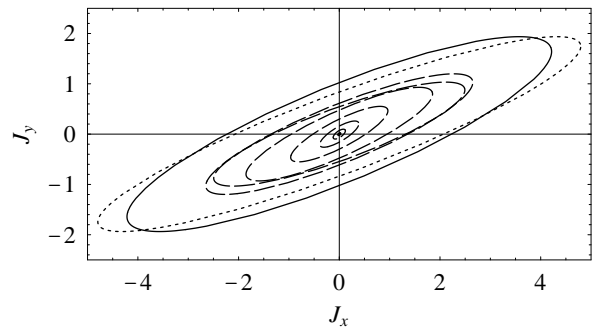}
  \caption{\label{fig:2}
    The distribution of the spin components orthogonal to the mean
    spin after performing a one-axis twisting squeeze on a system of
    $15$ spin $\frac{1}{2}$ particles with initial single particle
    density matrix given by \eqref{eq:rho_p} with $p=0.9$.  The
    squared length of the minor and major axes of the ellipse drawn
    with a solid line equal the variance of the spin-component in these
    directions. Additionally the direction of the minor axis correspond
    to the direction of minimal variance.
    The dotted ellipse shows, for comparison, the result of applying
    an optimal squeezing to an initially pure system ($p=1$), and the
    dashed ellipses illustrate the contributions from the various $j$
    subspaces in the mixed case as discussed in the text.}
\end{figure}

A one-axis twist about the $y$-axis, as mediated by the Hamiltonian
$\chi J_y^2$, will squeeze the variances of the $J_x$ and $J_y$ spin
components of an initially maximally polarized state as illustrated on
Fig.~\ref{fig:2} (the dotted ellipse): The variance of the spin
along one direction in the $xy$-plane is increased, while the variance
along the perpendicular direction is decreased to a value below the
standard quantum limit $j_0/2$.

As we have tried to illustrate by an example on the same figure, the
same is qualitatively true in the case where the initial state is not
a coherent spin state, but a ``sufficiently polarized'' mixed state.

Wave function components of definite $j$ show a reduced variance 
\begin{equation}
  \label{eq:varjtheta}
  \Delta J^2_\theta=\frac{j}{2}\left(
    1-\frac{2j-1}{4}\left[
      \sqrt{A_j^2+B_j^2}-A_j
    \right]
  \right) 
\end{equation}
of the spin component
$J_\theta = \cos \theta\, J_y - \sin \theta \, J_x$,
where $\theta=\frac{1}{2}\arctan \tfrac{B_j}{A_j}$, 
$A_j=1-\cos^{2j-2}\mu$, 
$B_j=4 \sin \tfrac{\mu}{2} \,\cos^{2j-2}\tfrac{\mu}{2}$ and
$\mu=2 \chi t$, \cite{KU93}.
We note that both the optimum interaction time and the direction
$\theta$ depend on $j$.
For a distribution of $j$-values this will deteriorate the squeezing.
Furthermore, the results in \cite{KU93} assume an initial state with
$m=j=j_0$, whereas we start out with a distribution of $m$-values.

Assuming that $p$ is so close to unity that we may use $j_c$ and
$\sigma$ from section \ref{sec:dens-matr-tens-1} to estimate the
center and width of the $j$-distribution, the relative variation of
the optimal $\mu$ over the width of the $j$-distribution is
approximately $\Delta \mu_0/\mu_0 \approx - \sqrt{(1-p)/j_0}$.
Consequently, for $j_0 \gg 1$ it is possible to choose a value of
$\mu$ that is nearly optimal for all populated $j$ subspaces.
According to \eqref{eq:rhojm} and \eqref{eq:margj}, we find that
within each $j$-subspace, the fraction of the population in the
uppermost $m$ level, $p^{(n)}_{j,j}$, is larger than $2-1/p$.

Similarly the direction of the squeeze axis does not seem to vary
significantly over the populated $j$ subspaces, as is also supported
by Fig.~\ref{fig:2} showing results for $15$ atoms with $p=0.9$.
Looking at the distributions within the different $j$-subspaces
(dashed ellipses), one should note that the outermost ellipse
corresponds not to the maximal $j$ but to $j=j_0-1$, which has the
greatest population. $j_0$ correspond to the second-largest ellipse
and we can see how the squeezing axis turns clockwise with increasing $j$.
When we perform the sum over all $j$-values we obtain the result
illustrated by the solid ellipse in Fig.~\ref{fig:2}.
The state is significantly squeezed, but less than in the ideal case
indicated by the dotted ellipse.
As anticipated, the population of the states with non-maximal $m$
within each subspace does not appear to have serious adverse effect on
the squeezing process.

\subsection{Two-axis countertwisting}
We analyze the time evolution under the two-axis countertwisting
Hamiltonian, $H_{ct}=\chi/2i \left(J^2_+-J^2_-\right)$.  One advantage
of the two-axis countertwisting Hamiltonian is that the squeezing is
always in the same direction. A disadvantage is that it cannot be
solved analytically, and we will proceed according to the
Holstein-Primakov approximation which is only valid for highly
polarized samples and not too strong squeezing.
Consequently we shall restrict the validity of our results to lowest
order in $1-p$ and assume $j_0$ to be much larger than unity.
Using results from \cite{MW95} we find the variance of 
the squeezed component, $J_y$, of $\vec{J}$ after squeezing to be 
\begin{equation}
  \label{eq:diagonalSqueeze}
  {\Delta J_y}^2= \sum_{j,m} \rho_{j,m} \frac{1}{2} \left(
    j(j+1)-m^2\right) e^{-\lambda j},
\end{equation}
where $\lambda=4 \chi t$.
In the fully polarized limit ($p=1$) we have 
${\Delta J_y}^2= \exp(-\lambda j_0)\,j_0/2$, 
and consequently  
$s_0=\exp(\lambda j_0/2)$ 
is the pure state squeezing factor.

It is a straightforward task to perform the sum over $m$
in \eqref{eq:diagonalSqueeze}. Doing this we obtain two terms inside
the $j$-sum, one of which may safely be dropped under the assumptions 
already made
($p>\frac{1}{2}$ and $j_0\gg 1$).
Keeping the remaining term we find
\begin{equation}\begin{split}
  \label{eq:uglysum}
  {\Delta J_y}^2 \approx &
  e^{-\lambda j_0}
  \left(\frac{p}{p_\lambda}\right)^{2 j_0} 
  \frac{p}{(2 p -1)^2}\\
  &\sum_{j} \frac{j^2}{j+j_0}
  \binom{2 j_0}{j_0+j}{p_\lambda}^{j_0+j}(1-p_\lambda)^{j_0-j},
\end{split}\end{equation}
where we have used the asymptotic approximation, $j_0\gg1$,
extensively and introduced
$p_\lambda$ as 
$p_\lambda=p/(1+(\exp(\lambda)-1)(1-p))$.
The three last terms of the second line are recognized as a normalized, rather
narrow, binomial distribution. Approximating this distribution with a
$\delta$-function centered at $j=(2 p_\lambda -1)j_0$ we find
\begin{equation}
  \label{eq:j2approx}
  {\Delta J_y}^2\approx
  e^{-\lambda j_0} \frac{j_0}{2}
  \left(\frac{p}{p_\lambda}\right)^{2 j_0+1} 
  \left(\frac{2 p_\lambda-1}{2 p -1}\right)^2 .
\end{equation}
To first order in $(1-p)$ this is equal to
\begin{equation*}
  {\Delta J_y}^2 \approx \frac{j_0}{2} e^{-\lambda j_0} 
  \left(1+(e^\lambda -1)
  (2 j_0 -3)
  (1-p)\right).
\end{equation*}

Based on this result we may determine the effect of the non-purity of
the initial distribution on $\xi_R$:  
the ratio between the frequency uncertainties obtainable by Ramsey
spectroscopy on squeezed and unsqueezed samples,
as introduced by Wineland {\it et.al.}~\cite{WB94}. According to this
reference we have
\begin{equation}
 \label{eq:squeezefact}
 \xi_R \equiv \frac{\Delta \omega}{\left(\Delta \omega\right)_0}
 =\frac{\Delta J_y / \left|\Braket{J_z}\right|}{
   \left(\Delta J_y / \left|\Braket{J_z}\right|\right)_0},
\end{equation}
where the terms in the denominator refer to the unsqueezed sample.
Since $\left(\Delta J_y^2\right)_0 = j_0/2$ is independent of $p$, and
$\Braket{J_z}=\left(\Braket{J_z}\right)_0$ to first order in $\lambda$, 
we find that to lowest order in $(1-p)$ and $\lambda$ we have
\begin{equation}
  \label{eq:xifinal}
  \xi_R \approx \frac{1}{s_0}
  \left(1+2(1-p)\log s_0 \right),
\end{equation}
that is: to first order in $1-p$ we get a correction to the quality of
the squeezed ensemble as compared to the case of performing the same
squeezing operation on an initially perfectly polarized ensemble.

Note that we have only examined how imperfect initial state
preparation affects the Ramsey spectroscopy; for an analysis of 
the significance of decoherence during the measurement see  Refs.
\cite{ulam-orgikh01:spin_squeez_decoh_limit_ramsey,huelga97:improv_frequen_stand_with_quant}.

\section{\label{sec:atoms-with-more}Atoms with more than two levels}
One of the virtues of the combinatorial analysis of collective spin
distributions is that it may be generalized to the case where the
atoms have spin $s>\frac{1}{2}$.

\subsection{Collective spin states for $n$ spin $s$ particles}
Proceeding as in section \ref{sec:coll-spin-stat} we find that the
dimension of the $\Ket{\Jm{j}{m}}$ subspace is
\begin{equation}
  D^{(s)}_j=\binom{n}{j_0+j}_{2s}-\binom{n}{j_0+j+1}_{2s}
\label{eq:SM_dimension2}
\end{equation}
where the generalized binomial $\binom{n}{m}_k$ denotes the number of
ways of placing $m$ indistinguishable balls in $n$ distinguishable
urns so that no urn contains more than $k$ balls. 
Note that this definition agrees with that of the ordinary binomial for $k=1$.

The main problem with this approach is that the generalized binomials
are not as analytically accessible as the binomials. Their
numerical values may, however, easily be obtained through the recursion
relation
\begin{equation}
  \label{eq:genbinrec}
  \binom{n}{m}_k = 
    \sum_{i=0}^k \binom{n-1}{m-i}_k,
\end{equation}
and their analytical properties are to some extent accessible
from the generating function:
\begin{align}
  \label{eq:boxgenerator}
  \sum_m \binom{n}{m}_k t^m 
  &= \left( 1+t+\ldots+t^k \right)^n\notag\\ 
  &= \left( \frac{1-t^{k+1}}{1-t} \right)^n.
\end{align}

\subsection{\label{sec:density-mixed-morethantwo}The 
density matrix for a tensor product of mixed states}
 
In order to represent the tensor product  $\rho_n = \rho_1^{\otimes n}$
in the basis of eigenstates of the collective angular momentum, we have 
to develop this basis and to identify the matrix elements of $\rho_n$.
It turns out to be advantageous to perform these two tasks in parallel
by sequential coupling of angular momenta of 1, 2,\ldots, $k$ spin $s$
particles, and determination of the appropriate density matrices 
$\rho_1,\rho_2,\ldots,\rho_k$. 

Let us introduce the collective spin of the first $k$ particles as
\begin{equation}
  \label{eq:def-subcoll}
  \vec{J}^{(k)} \equiv \sum_{i=1}^k \vec{S}_i,
\end{equation}
We denote by $\left\{\Ket{j_1,j_2,\ldots,j_k,m_k}\right\}$
the simultaneous eigenstates of 
$\bigr(\vec{J}^{(1)}\bigl)^2,\ldots,\bigl(\vec{J}^{(k)}\bigr)^2$
and $J^{(k)}_z$ with eigenvalues
$j_1(j_1+1),\ldots,j_k(j_k+1)$ and $m_k$ respectively.
These states are obtained by sequential coupling of the atomic spins:
by coupling the first, $j_1=s$, and second atom we obtain a total
spin $j_2$, which is subsequently coupled with the third atom to the
resulting spin $j_3$, etc.

The density matrix $\rho_1$ is our known input. 
We assume that $m_s$ labels the basis vectors in which $\rho_1$ is diagonal
with eigenvalues $p_{m_s}$ (note that $m_s$ does not need to be
associated with any real angular momentum or spin of the physical
system, the spin representation is convenient to represent any
$(2s+1)$-level atomic system).

If we assume that the density matrix, $\rho_k$, of $k$ atoms
is known in the basis of angular momentum coupled states,
$\left\{\Ket{j_1,j_2,\ldots,j_k,m_k}\right\}$,
we can recursively construct the density matrix of $k+1$ atoms:
 \begin{equation}
   \label{eq:notreduced}
   \begin{split}
    \rho_{k+1}
    = &\sum_{\alpha_k,\alpha_k',m_k,m_s}
    \Ket{\alpha_k',m_k}\Bra{\alpha_k,m_k}\otimes 
    \Ket{\Jm{s}{m_s}}\Bra{\Jm{s}{m_s}}\\
    &\centerBraket{\alpha_k',m_k}{\rho_k}{\alpha_k,m_k}
    p_{m_s},
  \end{split}
\end{equation}
where we have utilized that $\rho_k$ is diagonal in $m_k$ as
argued in section \ref{sec:dens-matr-tens-1}, and we have introduced 
$\alpha_k$ as a shorthand for the intermediate angular
momenta $j_1,j_2,\ldots,j_k$. 
It is now just a matter of angular momentum coupling to obtain from
(\ref{eq:notreduced}) the density matrix $\rho_{k+1}$ in the basis of
total angular momentum states of the $k+1$ particles, specifying
$j_{k+1}$, $m_{k+1}$ and all the intermediate angular momenta
$\alpha_k$.

The bookkeeping seems to be a demanding problem, but as in Eq.
\eqref{eq:rhored} we are only interested in the reduced density matrix
retaining the properties of the total spin of the $k+1$ particles.
Hence we trace over $\alpha_k$, {\it i.e.}, we add incoherently the
contributions from subspaces with different intermediate $j_{k'}$:
Let $p^{(k)}_{j_k,m_k}$ denote the diagonal elements of the reduced
density matrix $\rhored_k$, that is, the probability that $k$ atoms
have a total angular momentum of $j_k$ and projection $m_k$.
We can then write the matrix elements of the reduced density matrix for
$k+1$ atoms:
\begin{align}
  \label{eq:rho-red-general}
  &\largeBraket{\Jm{j_{k+1}'}{m_{k+1}'}}{\rhored_{k+1}}{\Jm{j_{k+1}}{m_{k+1}}} =\\ 
  &\sum_{j_k,m_k,m_s}
  \cgBraket{\Jm{j_{k+1}'}{m_{k+1}'}}{\Jmjm{j_k}{m_k}{s}{m_s}} 
  \cgBraket{\Jmjm{j_k}{m_k}{s}{m_s}}{\Jm{j_{k+1}}{m_{k+1}}}
  \,p^{(k)}_{j_k,m_k}\, p_{m_s},\notag
\end{align}
where the brackets,
$\smashcgBraket{\Jmjm{j_k}{m_k}{s}{m_s}}{\Jm{j_{k+1}}{m_{k+1}}}$,
denote Clebsch-Gordan coefficients.
An interesting consequence of the relation \eqref{eq:rho-red-general} 
is that for $k$ less than $n$ it is sufficient to determine recursively the
diagonal elements of the reduced density matrix, $\rhored_k$:
$p^{(k)}_{j_k,m_k}=\sum_{j_{k-1},m_{k-1},m_s}
\left|
\smashcgBraket{\Jmjm{j_{k-1}}{m_{k-1}}{s}{m_s}}{\Jm{j_k}{m_k}} 
\right|^2
\,p^{(k-1)}_{j_{k-1},m_{k-1}}\, p_{m_s}$.
In the last step we obtain $\rhored_n$ which also has coherences
between different collective spin states.  According to the selection
rules of angular momentum coupling this implies that the reduced
density matrix must be diagonal in $m_n$ and band diagonal in $j_n$
with values ranging over $| \Delta j_n |\le 2 s$.  

As an example, consider two spin $1$ particles which are both in the
pure state $\Ket{\Jm{1}{0}}$. In the collective spin basis the
combined state of these particles is
\begin{equation}
  \label{eq:nondiagex}
  \Ket{\Jm{1}{0}} \otimes \  \Ket{\Jm{1}{0}} =
  \sqrt{\tfrac{2}{3}}\, \Ket{\Jm{2}{0}}
  -\sqrt{\tfrac{1}{3}}\,\Ket{\Jm{0}{0}},
\end{equation}
which clearly yields a non-diagonal reduced density matrix.

Note that according to \eqref{eq:rhojm} we actually have the stronger
selection rule $\Delta j = 0 < 2 s$ for $s=\frac{1}{2}$.

In closing, we present a simple case of a mixed state of a system
of uncorrelated atoms with more than two levels, which might serve as an
approximation of {\it e.g.}, an optically pumped system.
We assume the single atom density matrix, $\rho_1$, to be diagonal, and of the
form
\begin{equation}
  \label{eq:thermaldist}
  p_{m_s} =
  \frac{1}{Z} e^{-\beta m_s},
\end{equation}
where $Z$ is a normalization constant depending on the parameter
$\beta$.
By induction through \eqref{eq:rho-red-general},
or by the 
argument that we used to deduce the $p$ dependence in section
\ref{sec:dens-matr-tens-1}, we find that with this form of $\rho_1$,
the parameter dependence of the reduced density matrix,
$\rhored_n$, must be $Z^{-n} \exp(-\beta m)$.
By the same arguments as in section \ref{sec:dens-matr-tens-1} we
conclude that the reduced density matrix is diagonal and
\begin{equation}
  \label{eq:rhothermal}
  p^{(n)}_{j,m} = D^{(s)}_j Z^{-n} e^{-\beta m},
\end{equation}
where $D^{(s)}_j$ is given by \eqref{eq:SM_dimension2}.

\section{\label{sec:conclusion}Summary}
In conclusion, we have reiterated the convenience of collective spin
operators for analyzing the behavior of a system of atoms.

We have established the explicit form of the reduced density matrices
of ensembles of two and multi-level atoms with a given single particle
distribution, and we have used these explicit distributions to predict
the effect of non-perfect sample preparation on the effectiveness of
spin squeezing by one-axis twisting and two-axis countertwisting.

\begin{acknowledgments}
This work was supported by the Danish Research Foundation 
--- Danmarks Grundforskningsfond and by the Information Society
Technologies Programme IST-2000-30064, REQC.
\end{acknowledgments}

%\bibliography{colspin}
% And the insert the .bbl file below:

\end{document}